\documentclass [12pt,twoside]{article}           

\usepackage{epsfig,times,lscape}
\usepackage[usenames]{color}

    \ifnum\count102=1

    \fi

    \ifnum\count102=2
\fi

\newcommand{\nc}{\newcommand}

     \ifnum\count101=1
\nc{\qI}[1]{\section{{#1}}}
\nc{\qA}[1]{\subsection{{#1}}}
\nc{\qun}[1]{\subsubsection{{#1}}}
\nc{\qa}[1]{\paragraph{{#1}}}

\def\qpar{\vskip 2mm plus 0.2mm minus 0.2mm}
\def\qL{\hfill \break}
     \fi 

      \ifnum\count101=2
 \nc{\qI}[1]{\parindent=0mm \vskip 8mm 
{\centerline{\LARGE \color{red}#1}}\vskip 3mm}
\nc{\qA}[1]{\vskip 2.5mm \noindent 
{{\bf\large\color{blue}  #1}} \vskip 1mm \parindent=0mm}
 \nc{\qun}[1]{\vskip 1mm \noindent {\sl #1 }\quad }

\def\qL{\hfill \break}
\def\qpar{\vskip 2mm plus 0.2mm minus 0.2mm}

      \fi

\nc{\qfoot}[1]{\footnote{{#1}}}

      \ifnum\count101=1
\def\qbu{\hfill \par \hskip 6mm $ \bullet $ \hskip 2mm}
\def\qee#1{\hfill \par \hskip 6mm (#1) \hskip 2 mm}
      \fi
      \ifnum\count101=2
\def\qbu{\hfill \par \hskip 4mm $ \bullet $ \hskip 2mm}
\def\qee#1{\hfill \par \hskip 4mm (#1) \hskip 2 mm}
      \fi

\def\qparr{ \vskip 1.0mm plus 0.2mm minus 0.2mm \hangindent=10mm
\hangafter=1}

     \ifnum\count101=1 
 \def\qdec#1{\parindent=0mm\par {\leftskip=2cm {#1} \par}}
     \fi
     \ifnum\count101=2
  \def\qdec#1{\parindent=0mm \par {\leftskip=1cm {#1} \par}}
  
  \def\qcitb#1{\noindent \hbox to 102mm{\hfill \small #1} \vskip 1mm}
      \fi

 \def\qpages#1{\count102=0{\loop\advance\count102 by 1
 \null \vfill\eject \ifnum\count102<#1 \repeat}}

\def\qn#1{\eqno \hbox{(#1)}}

\def\qv{\vskip 0.1mm plus 0.05mm minus 0.05mm}

\def\qhw{\hskip 1.5mm}
\def\qleg#1#2#3{\noindent {\bf \small #1\qhw}{\small #2\qhw}{\it \small #3}\qv }

\begin{document}
\thispagestyle{empty}



\markboth{{\sl \hfill  \hfill \protect\phantom{3}}}
        {{\protect\phantom{3}\sl \hfill  \hfill}}

\color{yellow} 
\hrule height 20mm depth 10mm width 170mm 
\color{black}
\vskip -2.2cm 
\centerline{\bf \Large Extending physical chemistry}
\vskip 2mm
\centerline{\bf \Large to populations of living organisms.}
\vskip 2mm
\centerline{\bf \Large First step: measuring coupling strength}
\vskip 4mm
\centerline{\large 
Zengru Di$ ^1 $,
Bertrand M. Roehner$ ^2 $
}


\vskip 8mm
\normalsize
{\bf Abstract}\quad For any system, whether physical or non-physical,
knowledge of the form and strength of inter-individual
interactions is a key-information. In an approach based on
statistical physics one needs to know the interaction in order
to write the Hamiltonian of the
system: $ H=H_{\hbox{free}}+H_{\hbox{interaction}} $. For non-physical
systems, based on qualitative arguments similar to those used
in physical chemistry, interaction strength
gives useful clues about the macroscopic properties of the system
(e.g. for an institution the dropout rate is expected
to be smaller when the inter-individual attraction is stronger).\qL
Even though our ultimate
objective is the understanding of social phenomena, we found that
systems composed
of insects (or other living organisms) are of 
great convenience for investigating {\it group effects}.
In this paper we show how to design experiments that enable us
to estimate the strength of interaction in groups of insects.
By repeating the same experiments
with increasing numbers of insects, ranging from less than 10 to several
hundreds, one is able to explore key-properties
of the interaction. \qL
The data turn out to be consistent with 
a global correlation that is independent of distance 
(at least within a range
of a few centimeters). Estimates of this average cross-correlation
will be given for ants, beetles and fruit flies. 
The experimental results
clearly exclude an Ising-like interaction, 
that is to say one that would
be restricted to nearest neighbors. 
In the case of fruit flies
the average cross-correlation appears to be negative which means
that instead of an inter-individual attraction there is a 
(weak) repulsive effect.\qL
In our conclusion we insist on the fact that such ``physics-like
experiments'' on insect populations
provide a valuable alternative
to computer simulations. When testable group effects are predicted 
by a model,
the required experiments can be set up {\it within a short time},
thus permitting to confirm or disprove the model.
This marks
a significant progress with respect to modeling of social systems where,
all too often, the requested statistical data just do not
exist, thus obstructing any
fruitful dialogue between theory and observation. 

\vskip 3mm
\centerline{\it First version: 20 December 2012}

\vskip 3mm
{\normalsize Key-words: Group effects, collective behavior,
physical chemistry, statistical physics,
living organisms, insects, ants, drosophila,
coupling strength, inter-correlation, 
correlated random variables, exchangeable random variables.}
\vskip 2mm

PACS classification: Interdisciplinary (89.20.-a) +
Correlation, collective effects (71.45.Gm)
\vskip 5mm

{\normalsize 
1: Department of Systems Science, Beijing Normal University, 
Beijing, China.\qL
Email: zdi@bnu.edu.cn \qL
2: Institute for Theoretical and High Energy Physics (LPTHE),
University Pierre and Marie Curie, Paris, France. \qL
Email: roehner@lpthe.jussieu.fr
}

\vfill\eject

\large

In a first version of the paper the title made reference
to ``statistical physics'' rather than to ``physical
chemistry''. Although chemistry itself plays no role in our
investigation, we think it is important to emphasize that
at this point it relies rather on
the approach of ``physical chemistry'' than on that of
statistical mechanics. The reasons for that will be
explained later on%
\qfoot{A preliminary but extended version (some 100 pages) of 
the present paper is available on the following website:\qL
http://www.lpthe.jussieu.fr/~roehner/effusion.pdf}%
.

\qI{Rationale and motivations}

By using the theoretical framework of statistical mechanics
one can derive the macroscopic properties of a system
from the characteristics of its microscopic elements. This is a
major achievement and so it is hardly surprising that researchers
from other disciplines (e.g. biology, demography, sociology or
economics) have been tempted to adapt such a powerful tool
to their own field. In light of the successful record of 
statistical mechanics in physics 
there is little doubt that 
such extensions appear highly desirable. 
Yet, to our best knowledge,
in spite of many attempts in this direction  
such attempts have not been highly successful so far%
\qfoot{Recently, some promising breakthroughs were made
in this direction by Japanese economists; see Aoki 
and Yoshikawa (2007), Iyetomi  et al. (2011), Iyetomi (2012).}%
.

\qA{Obstacles}
As a matter of fact this is hardly surprising for there
are indeed many obstacles.
\qbu Statistical physics is fundamentally a theory of
systems in equilibrium. For systems which are (strongly)
out of equilibrium the very concept of temperature
becomes meaningless.
\qbu Statistical physics relies on the identification
of ensemble averages (which are predicted theoretically)
and time-averages (which are measured in experiments).
This so-called ergodic hypothesis may be valid for 
physical systems which move
from one state to another every picosecond so that there
are trillions of transitions during an observation time
of a few seconds. Yet, it is not obvious that 
such an assumption can still be accepted for
socio-economic systems for which the transition rates
are much slower%
\qfoot{The highest transition rates are probably those
in currency exchange markets with hundreds of orders (worldwide)
every second. Recently so-called  high speed trading,
that is to say transaction orders passed by computers,
has reduced transition times to a few micro-seconds
at least for a number of actively traded securities.}%
.
\qbu Last but not least, one should not forget that in
order to use the theoretical framework of statistical
mechanics one needs to know the Hamiltonian $ H $ of the system
which indicates how energy is distributed in the system.
Generally $ H $ includes three parts:
$$ H=H_0 +H_{\hbox{inter}} + H_{\hbox{exo}} $$
where $ H_0 $ stands for the free particles,
 $ H_{\hbox{inter}} $ for the interaction energy between them
and $ H_{\hbox{exo}} $ for the energy of the particles when
an exogenous field is involved. For instance
$ H_0\sim \sum p_i^2/m $ for a system containing the molecules of
a gas, $ H_{\hbox{inter}} \sim \sum{ 1\over (r_i-r_j)^6 } $
when one wants to take into account the van der Waals forces
between the molecules, and $ H_{\hbox{exo}}\sim \sum S_iH(i) $
for the energy of a set of spins in an external magnetic
field. 
\qpar
Whereas the third term can possibly be omitted when the
experimental device can be shielded from external fields,
the interaction
term must {\it always} be taken into account%
\qfoot{Even in order to use a mean field approximation
one must know the form of $ H_{\hbox{inter}} $.}%
.
Needless to say, there are almost
no biological or social systems for which one has a clear
knowledge of their interactions. 
It is precisely the main purpose of the present paper to 
explain how such interactions can be measured. 

\qA{Reasons for optimism}
The previous list of obstacles could appear discouraging
especially if one realizes that there are many other
problems in non-physical systems
just for defining key-variables such as 
velocity or energy. However, there are also good reasons for
optimism as we will see now.
\qpar
First it can be observed that the theory of phase transitions 
has been used to describe the transition between ordinary
hadronic matter and quark-gluon plasma. As such states
are characterized by temperature of the order of $ 10^{12} $K
and life-times of the order of $ 10^{-20} $s, it means
that this theory is applied well beyond the limits
of the phenomena%
\qfoot{E.g. second order
phase transitions such as the paramagnetic-ferromagnetic transition
in iron.} 
for which it was originally developed. 
Does the ergodic assumption hold
for such extremely short time intervals? Nobody knows and probably
nobody cares. The strategy of physicists is to use this framework
without giving too much concern to underlying assumptions. 
If sensible results emerge this will provide so to say
{\it ex post} justification.
\qpar
Secondly, it can be observed that the title of this paper
does not refer to statistical physics but to physical
chemistry. Why?
\qbu Although the objective of physical chemistry is also
to explain the properties of macroscopic systems in terms
of molecular interactions, there are two main differences
with the approach of statistical mechanics. 
First, physical chemistry considers a broad range 
of molecules rather than just the simplest ones as is
done in physics. Thus, because many cases are being
considered, it becomes indispensable to adopt a comparative 
perspective. Why is the melting point of argon lower
than the melting point of water? Why is the equilibrium
vapor pressure higher for ethanol than for water? 
And so on and so forth. 
\qbu Because it would be an almost impossible task
to propose (and solve)
full-fledged models for all these cases, physical chemistry 
will rather resort to qualitative arguments.
For instance, a standard argument is to observe that
the stronger are molecular interactions in a liquid, the 
fewer molecules will be able to escape which in turn
will lead to a low equilibrium vapor pressure above
the liquid. Whereas this argument relies on a 
specific mechanism describing how molecules leave the
liquid, it does not require any of the assumptions
that we listed in previously. Even
equilibrium is not strictly required. Indeed, if the
container is left open, no equilibrium will take place
and no equilibrium vapor pressure can be defined,
but the same argument can nevertheless be used 
for explaining differences in the evaporation rate.
\qpar

Such kind of argument can be used with success to
explain many physical properties. For instance, the boiling
temperature of alkanes 
($ \hbox{C}_{\hbox{\small n}}\hbox{H}_{\hbox{\small 2n+2}} $)
is expected to
increase with $ n $ because the so-called London 
attraction forces (due to induced polarization which
create short-lived dipoles that attract one another)
exist between {\it all} atoms and therefore, 
in the absence of any other force,
attraction will be stronger for big molecules
than for small ones. Through a similar argument one would
also expect the heat of vaporization to increase with $ n $.  
These predictions are indeed confirmed by experimental
data; two graphs displaying such data can be found
in Roehner (2004, p. 663).
\qpar

In short, once one knows the strength of interaction
in a system, one should be able to derive several of its 
macroscopic properties. Thus, we are again confronted to
the same key-question: {\it how can we measure interaction 
strengths?} To answer this question we will make yet
another simplification.
\qee{3} The simplifications that we have already made consisted
firstly in saying that we do not need to care too much about the
underlying hypotheses of statistical physics, secondly
that (at least in a first stage) there is no need to
use the mathematical framework of statistical mechanics. 
Now our objective has become less ambitious and 
the only question on which one needs to focus is to
develop {\it experimental} ways for measuring coupling strength
between the elements of the system. The word {\it
experimental}  leads us to a third simplification. 
\qpar

It is often said that for socio-economic systems one
cannot make experiments%
\qfoot{In the discussion which follows we leave apart so-called
class-room experiments that are performed with small
groups of students. Such experiments can be useful
to study how people will react in specific circumstances
such as in response to auction rules for instance.
However, one does not see how collective behavior
can truly be studied in such a way because the experiment
will only reflect genuine behavior if the people are {\it not}
told that they are involved in an experiment.
In the 1970s and 1980s the psycho-sociologist
Stanley Milgram has performed experiments of this kind.
However, such an approach raises major ethical problems
and should rather be avoided.}%
.
However, this is only partially true. In fact, social
sciences researchers are in the same position as astrophysicists.
While they cannot
perform {\it any} observation that they would like to do, 
nevertheless they can use such
statistical data that are available to make a limited number of
observations%
\qfoot{Researchers who have appropriate funding can
even organize surveys in order to collect data that would not
be available otherwise.}%
.
Yet, one must recognize that in many investigations
the very data that one would need turn out to be unavailable.
This is a serious obstacle. The task of designing appropriate
measurement methods is difficult enough in itself; it would
become altogether impossible if at each step 
progress is hindered by a lack of data. 
\qpar

There is a simple solution. Instead of studying people we
can study populations of living organisms such as 
bacteria, insects or small fishes. 
For all these populations there exists
a broad range of species. Different species will have different 
inter-individual interactions. Thus, one is very much in the
same position as in physical chemistry. In what follows we will
limit ourselves to populations of insects. 
\qpar

Our goal is to study groups of insects not at all as
an entomologist would do but from the
perspective of physical chemistry. In this respect 
living organisms have another important advantage over social
or economic systems. Energy is a key-notion in physics. 
While it is not obvious how to define 
the ``energy'' of a set of stocks
or a sample of companies, it is  easy
to define the velocity and kinetic energy of a group of ants. 
In other words, systems of living organisms are much closer
to physical systems than are socio-economic systems. 
\qpar

In the next section we explain how we designed and implemented
our experiments. 
In the last section we propose some consistency tests of
our results.

\qI{Design of the experiments}

The experiment will be described for ants but their design 
is fairly similar for other insects such as 
fruit flies or beetles.
\qpar
A number $ n $ of 
ants are contained in a rectangular box (15cm long and
5cm wide, 4mm high)  (see Fig. 1a). In this box one defines
two part: an area $ A $ and the part $ B $ of the box which
does not belong to $ A $. For the sake of simplicity
we can think of $ A $ as being the left-hand side of the box
as is the case in Fig. 1a. However, one should keep in mind that
$ A $ can also be much smaller than one half of the box.
This allows to explore the behavior of the ants at smaller scale.
\qpar

The ants can choose the compartment in which they wish to go
or to stay. We record the number $ n_A(t) $
of ants which are in compartment $ A $ at time $ t $.
\qpar
The idea of the experiment is the following.
\qpar
\qbu Suppose for a moment that the movements of the ants are
completely correlated. This means that if one ant goes
from $ A $ to $ B $ (or from $ B $ to $ A $)
all the others will follow. Thus at each time step
$ n_A(t) $ may experience huge jumps, either from $ N $ to $ 0 $ 
or vice versa. 
\qbu Suppose now that there is  a zero correlation  between
the movements of the ants. This means that if one ant goes
from $ A $ to $ B $ it will not be imitated by others.
Of course, other ants may make the same move but they will
do so independently from one another. As a result their
moves will follow a binomial process. A move of all the 
ants together is not completely excluded but it will
occur with a probability of $ (1/2)^N $ and decrease exponentially 
when $ N $ increases.
\qpar

\begin{minipage}[t]{0.4\textwidth}
\centerline{\psfig{width=4cm,figure=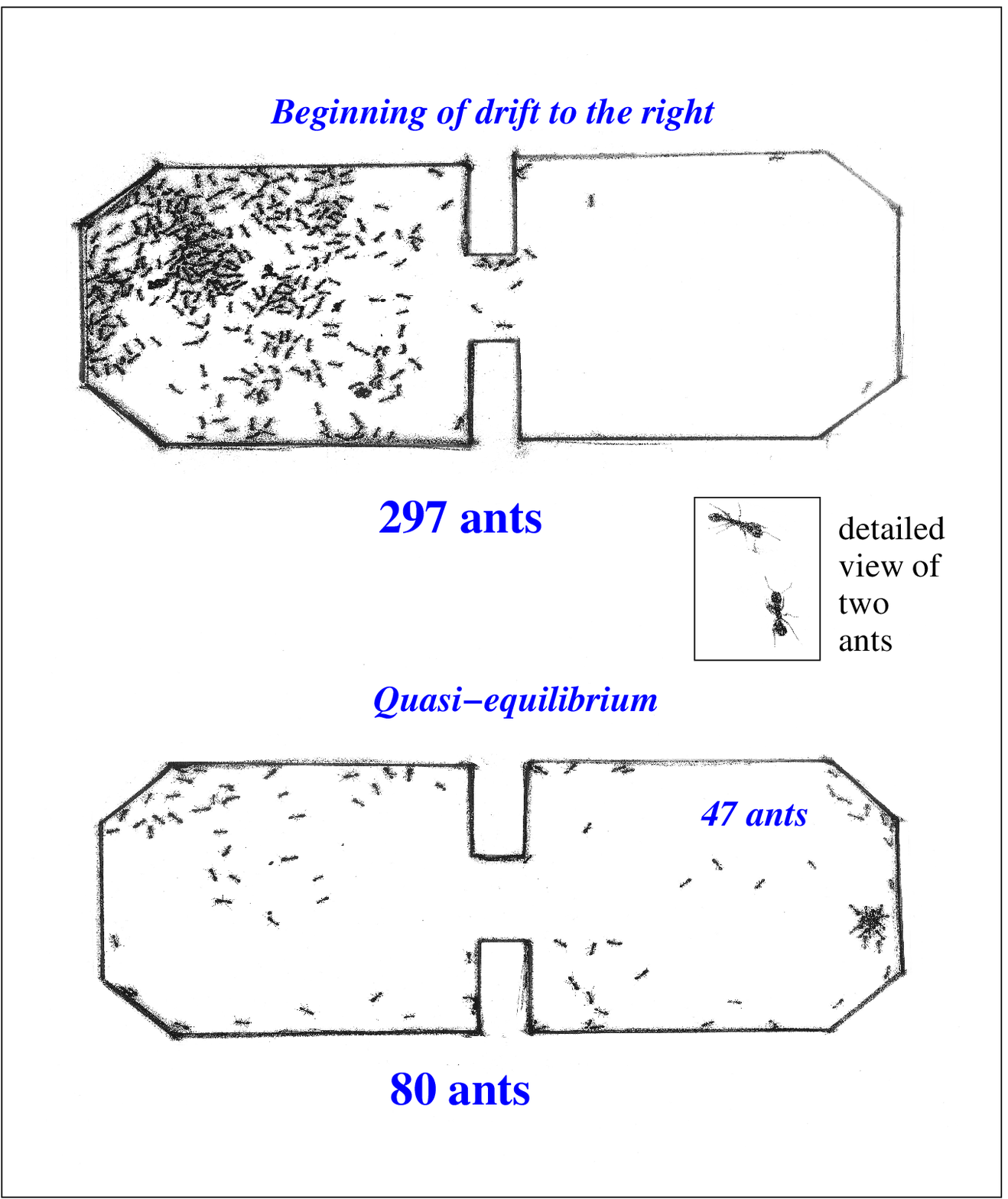}}
\end{minipage}
\begin{minipage}[t]{0.4\textwidth}
\null
\vskip -50mm
\centerline{\psfig{width=5.5cm,figure=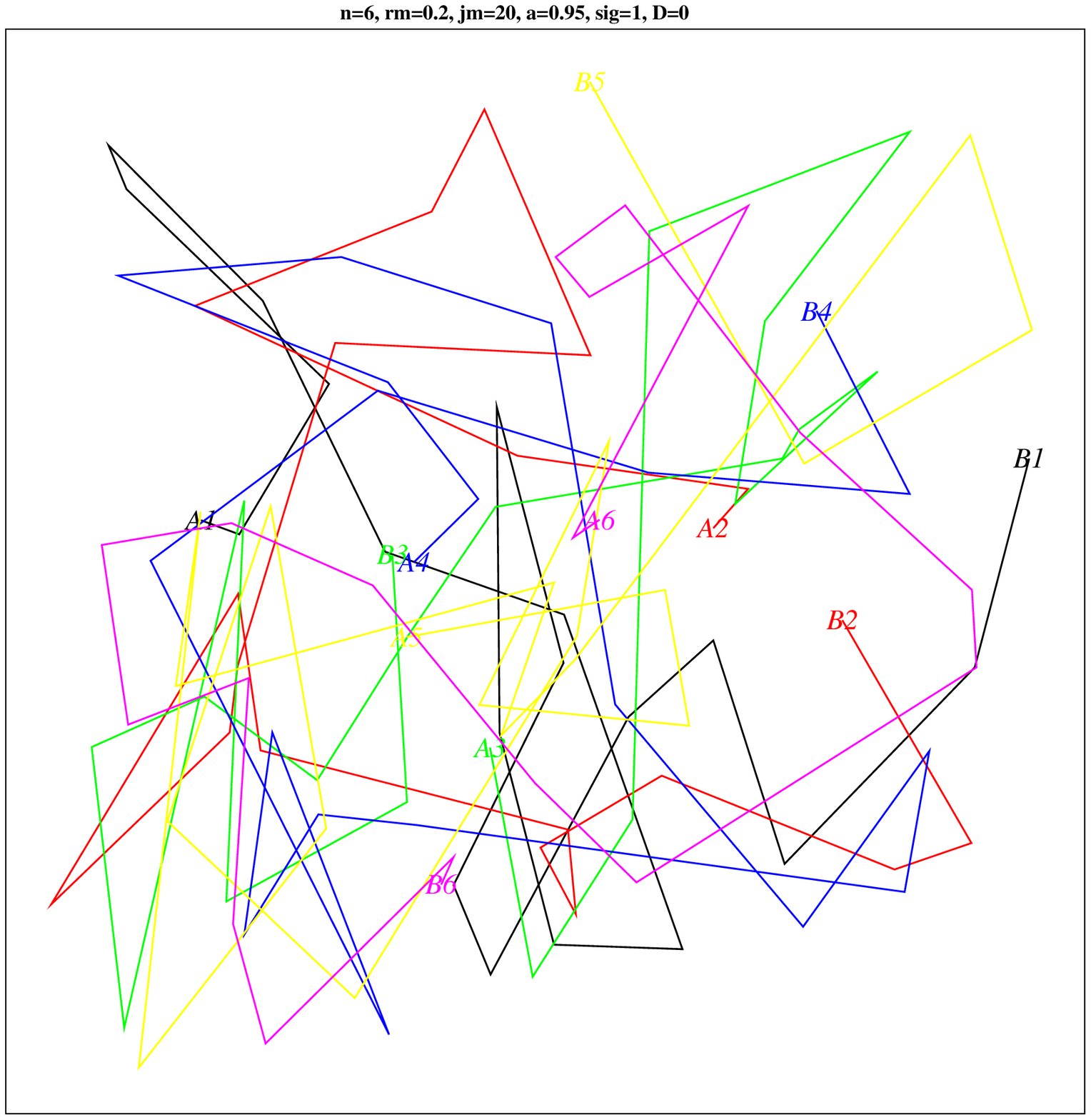}}
\vskip -5mm
\vskip 5mm
\end{minipage}
\vskip 2mm
{\color{blue} \small Fig. 1a (left): Experiments with ants in a 
two-compartment device. Fig. 1b (right): Simulation with 
an inter-individual cross-correlation equal to 0.2.}

This argument suggests that there is a connection
between the standard deviation of $ n_A(t) $ and the average
correlation between the movements of the ants.  
Needless to say, we wish to know the mathematical form
of this relationship. Then, by recording the fluctuations
of $ n_A(t) $ we will be able to compute its variance
and to derive the average correlation between ants. 
This average correlation can be considered as a measure
of their interaction strength. 
\qpar

\qI{Formalization}

To each ant $ i $ we associate a random variable $ X_i $
which takes the value $ 1 $ when $ i $ is in compartment 
$ A $ and $ 0 $ otherwise. Thus, at any moment t, 
the number of ants in compartment
$ A $ will be given by: $ S_n=\sum_1^n X_1 $.
If $ n_A(t) $ is a {\it stationary} random function, 
it is reasonable%
\qfoot{While of course necessary,
the stationarity condition is not sufficient
to guaranty ergodicity of the standard deviation. The 
specific mathematical condition that $ n_A(t) $ must satisfy is
given in Papoulis (1965, p. 330).}
to assume that the variance
computed from the time series $ n_A(t) $ coincides with
the probabilistic (i.e. ensemble) variance of the 
random variable $ S_n $. 
\qpar

Various assumptions can be made regarding inter-individual
interaction. Each assumption leads to different
correlations between the $ X_i $. We will examine
two specific cases: uniform correlation which means that
$ r_{ij}=r(X_i,X_j) $ is basically the same for all pairs $ (i,j) $
and correlations which decrease exponentially when the
difference $ i-j $ increases: $ r_{ij}\sim \eta^{|i-j|} $

\qA{Uniform correlation}

In this case, $ \sigma^2(S_n)=\sigma^2(n_A) $ is given by the following 
proposition.

\qdec{ {\color{blue} \bf Variance
of a sum of uniformly correlated variables.}\quad 
We consider a sum $ S_n $ of $ n $ identically distributed random
variables $ X_i $ of variance $ \sigma^2 $.
We assume that between $ X_i, X_j , i\not= j $ there are
cross-correlations $ r_{ij} $, the average of which is 
is denoted by $ \overline r $: 
$ \overline r = { 1 \over [n(n-1)/2] } \sum_{i<j}^n r_{ij} $.
Then, the variance of $ S_n=X_1+\ldots + X_n $ is given by:
$$ { \sigma^2(S_n)\over n\sigma^2 }= (n-1)\overline r +1 \qn{1} $$
}

The proof is fairly straightforward and is given in Appendix A. 
\qpar
Four observations are of interest in relation with formula (1).
\qee{1} The factor $ n\sigma^2 $ represents the variance of $ S_n $
when the variables are uncorrelated. Therefore the ratio
on the left-hand side represents the variance of $ S_n $ divided
by what it would be if the correlations are switched off. 
Subsequently this ratio will be denoted by $ g^2 $.
\qee{2}
In the special case where $ \overline r =1 $, formula (1)
gives: $ \sigma^2(S_n)=n^2\sigma^2 $. This result can be 
confirmed by observing that $ \overline r =1 $ means that all
variables $ X_i $ are identical that is to say take the same
values (with probability 1). Thus, 
$ S_n=nX_1 \Rightarrow \sigma^2(S_n)=\sigma^2(nX_1)=n^2\sigma^2 $
\qee{3} A negative average correlation reduces the variance
instead of increasing it. 
This would correspond to a repulsive force between the individuals.
It is of interest to observe that $ \overline r $ cannot
become smaller than $ -1/(n-1) $. In this case the variance is reduced
to zero. Intuitively, this corresponds to a situation where the
move of any individual is countered by the moves of the 
others in a way which leaves $ S_n $ unchanged. 
\qee{4} Formula (1) applies to any random variables $ X_i $.
For the problem in which we are interested, the $ X_i $
have a special meaning from which results that: 
$$ \sigma^2(X_i)=E(X_i^2)-E^2(X_i)=P\{X_i=1\}1-\left(P\{X_i=1\}1\right)^2 
=p(1-p) $$
where $ p $ is the fraction of $ A $ with respect to the total area. 

\qA{Ising-like correlations}
When the interaction is restricted to nearest neighbors as in the
one-dimensional Ising model for spins, the correlation between
the $ X_i $ decreases exponentially
when the ``distance'' between the spins
increases (Glauber 1963 p. 300). In this case the variance of 
$ S_n $ is given by the following proposition.

\qdec{ {\color{blue} \bf Variance
of a sum of Ising-like correlated variables.}\quad 
We consider a sum $ S_n $ of $ n $ identically distributed random
variables $ X_i $ of variance $ \sigma^2 $.
We assume that between $ X_i, X_j , i\not= j $ there is a 
cross-correlation $ r_{ij}=\eta^{|j-i||},\ 0<\eta<1 $. 
Then, the variance of $ S_n=X_1+\ldots + X_n $ is given by:
$$ { \sigma^2(S_n)\over n\sigma^2 }= 
{ 1+\eta\over 1-\eta } - { 2\eta \over n(1-\eta)^2 } \qn{2} $$
}

The proof is fairly similar to the proof of the first proposition
and it is outlined in Appendix A. 
According to this result, the ratio
$ g^2(n)={ \sigma^2(S_n)\over n\sigma^2 } $
is slightly increasing when $ n $ increases (see Appendix A).
However, when $ n $ becomes large the term
involving $ n $ becomes negligible with respect to the
first term. Thus, it is 
legitimate to say that for large $ n $, $ g^2(n) $ is almost constant.
\qpar

Can one explain the difference between case 1 and 2
intuitively? We have already 
observed that if $ \overline r $ is close to 1, almost all insects will
cross from one side to the other at the same time which will
result in big fluctuations of $ n_A(t) $ between $ 0 $ and $ n $.
In the second model the parallel of such a high correlation
would be $ \eta $ close to 1, e.g. $ \eta=0.9 $. Yet, even
with such a value of $ \eta $ the correlation between
$ i $ and its neighbors will fall off rapidly when the distance
increases.
This means that when 
$ i $ will change side, only a small number ($ f $) of neighbors 
will follow. As $ f $ depends only upon $ \eta $ (and not
upon $ n $) one sees that $ g^2(n) $ does not increase
with $ n $.
\qpar

In short, for the models that we considered  
the ratio $ g^2(n) $ can 
behave in three different ways as a function of $ n $. 
 \qee{1} It decreases linearly when $ \overline r<0 $
 \qee{2} It is almost constant when $ r_{ij} $ 
decreases exponentially with respect to $ |i-j| $.
 \qee{3} It increases linearly when $ \overline r>0 $. 
\qpar
We will see that only cases 1 and 3 occur in our observations.

\qI{Experimental results}

\qA{Procedure}
The experimental procedure involves the following steps.
\qbu First one must spread $ n $ ants fairly 
uniformly in the whole container.
Then pictures will be taken every $ \delta $ seconds
\qfoot{An ``appropriate'' time interval is important
for the accuracy of the measurement. Of course, it
is useless to take pictures when nothing happens
that is to say when $ n_A(t) $ does not change.
On the other hand, simulations show that one
can greatly improve the accuracy of the measurement by
increasing the number of pictures. In our experiments,
depending on the activity and number of insects
$ \delta $ was between 10s to 120s.}%
.
\qbu These pictures will allow us to record the numbers $ n_A(t) $. 
Once the variance of this time series has been computed one gets
the ratio $ g^2(n) $. 
\qbu
By repeating this procedure for different 
number of ants one gets results which can be represented as 
a set of points $ \left( n-1,g^2(n) \right) $ (see  Fig. 2).
\qbu
A linear regression performed on this set of points gives an estimate
of the slope $ \overline r $. 

\qA{Results} 
See the graphs in Fig. 2a,b,c.
\qpar
An important observation is in order regarding the magnitude of the
estimated average correlation. First, it must be emphasized that
$ \overline r $ is very different from the correlation estimated
from a scatter-plot. In the latter case a correlation as low as
0.01 would be non-significant (in the sense that the confidence
interval would contain 0) except if the scatter-plot contains
several thousand data points. Here, however, the correlation
was obtained as the slope of a regression line and its
estimate is quite significant as can be seen from the size of
the error bars. 
\qpar
In order to get an intuitive understanding of $ \overline r $,
one should compare the actual trajectories of the insects to
those shown in the simulation of Fig. 1b. Broadly speaking, the 
comparison will reveal that at individual level
the actual trajectories of the insects are even more random
than those in Fig. 1b. In spite of this high degree of randomness
there is an observable effect at the macro level. 
The situation is somewhat the same as for a gas or a liquid.
In spite of the randomness of the movements of individual
molecules there are nevertheless well defined macroscopic
properties.

\begin{figure}[htb]
\centerline{\psfig{width=7cm,figure=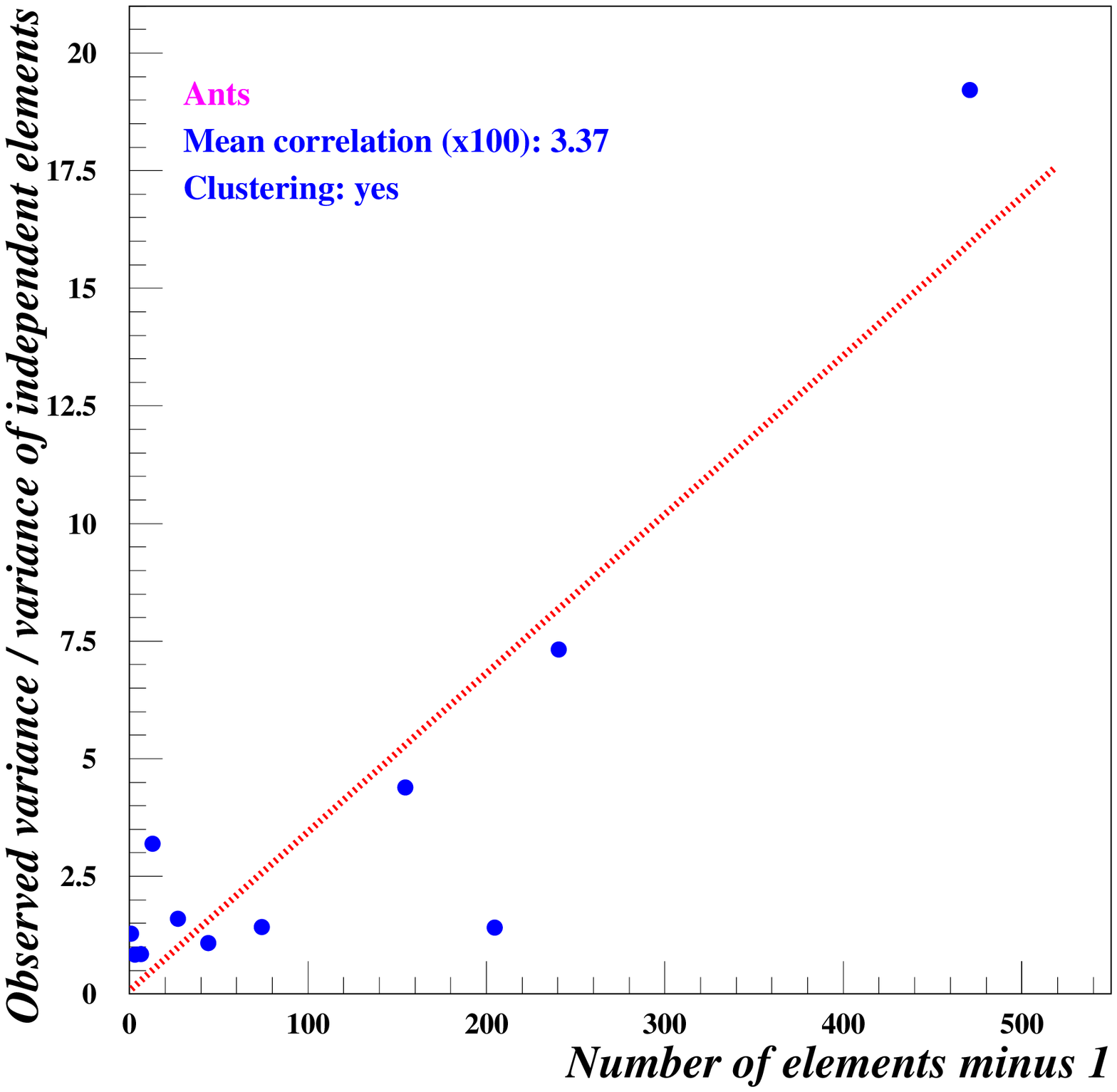}}
\vskip -2mm
\centerline{\psfig{width=13cm,figure=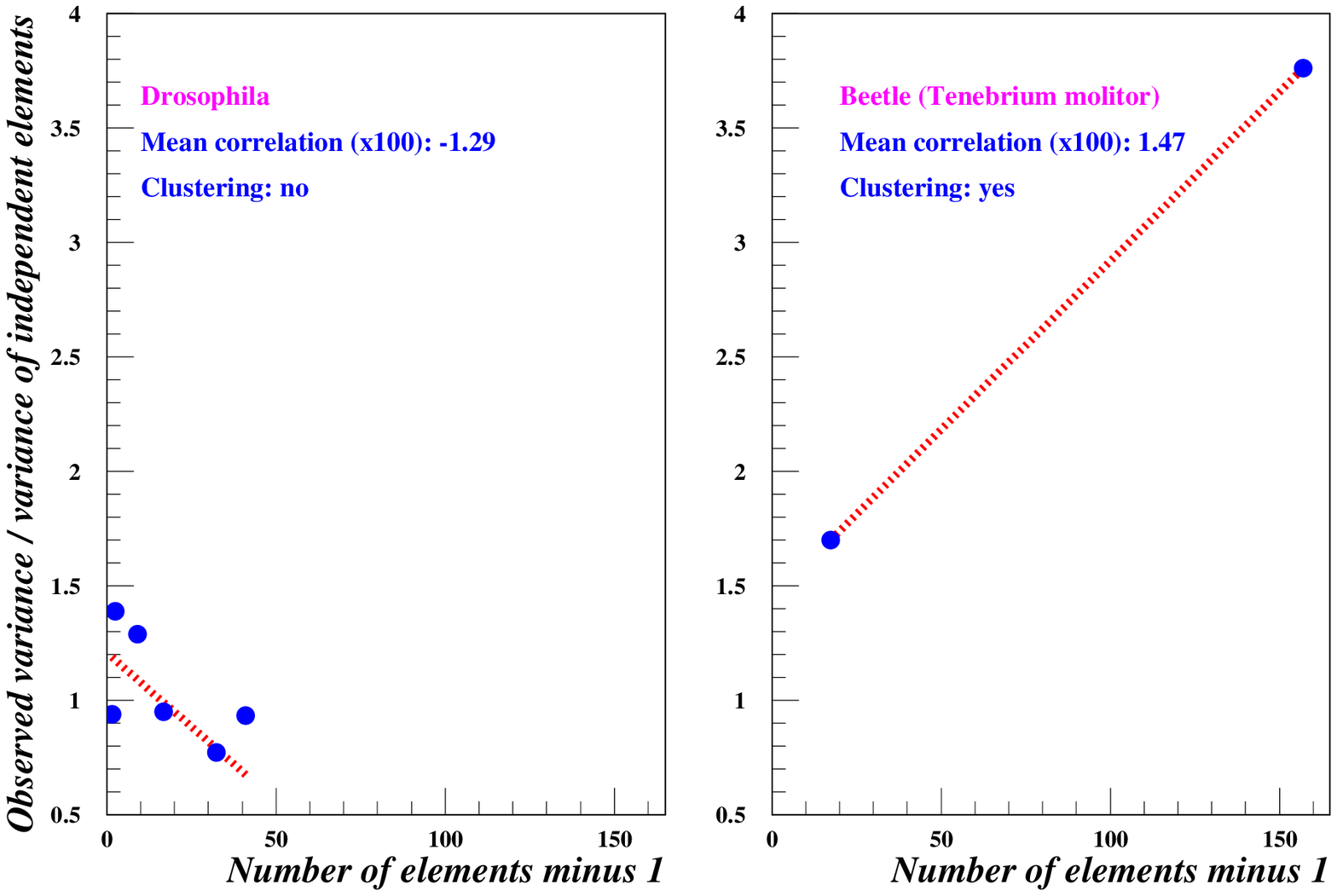}}
\vskip 1mm
\qleg{Fig. 2a,b,c: Relation between 
the variance of the number of individuals in
a compartment $ A $ and the size of the group.}
{We suppose that the whole domain which contains the
ants has been divided
into two parts and we observe the fluctuations $ n_A(t) $
of the number of individuals in part $ A $. 
The slope of the regression line gives an estimate of
the mean correlation between the moves of individual elements.
The negative correlation observed for drosophila
can be interpreted as the result of repulsive forces
between individuals.
The ability to form clusters can be seen as revealing the 
existence of attractive inter-individual forces. Thus,
this characteristics comes as a confirmation of the 
sign of the correlation.\qL
The confidence intervals (at a probability level of 0.90) are
as follows: ants: $ 100\overline r =3.37\pm 0.9 $, 
drosophila: $ 100\overline r =-1.29\pm 1.16 $, for the beetle
graph there are too few points to compute the confidence interval. 
\qL}
{These experiments were performed between June and October 2012
in three different places, first in
Paris (ants and drosophila), then in 
Beijing (drosophila) and finally in  Kunming, Yunnan 
Province, China (beetles).}
\end{figure}

\qA{Problems}
Although the procedure may appear fairly straightforward there
are a number of hurdles; while some are purely technical  
others are of more fundamental importance. Let us begin
with the latter.
\qpar
Ideally, in order to remain in a stationary equilibrium situation
one would like $ n_A(t) $ to fluctuate around $ 1/2 $. Actually,
for ants as well as for beetles, $ n_A(t) $ can become very 
different from 1/2. This is due to the fact that in such
cases the individuals will form a big cluster in one part
of the container. Thus, if the cluster is in $ A $, 
the ratio $ n_A(t)/n $
will become close to 1, whereas it will decrease toward $ 0 $
if the cluster is in part $ B $. 
\qpar
In a sense, this clustering behavior is good news because 
it is a direct proof of the existence of an 
inter-individual attraction. On the other hand, however, it
introduces a bias in the measurement of $ \overline r $. 
A correction procedure was introduced to take this effect into
account.

\qpar
There is a problem which arises especially for drosophila, namely
the fact that once introduced in the observation device
only a few of the insects will move. In the case of drosophila
this may take the following form: in a group of some 50 only 
about 5 to 10
will move at one moment and they
will do so with great speed going
from one end of the container to the other
without seemingly caring about the 45 others%
\qfoot{Whereas ants will tend to slow down or
stop every time they come close
to another ant.}%
. 
Another circumstance which will prevent the insects from
moving is when they form a cluster.
Although a correction
can be introduced in the analysis to take into account such ``frozen'' 
elements, it is clear that the analysis eventually
becomes meaningless when 
the proportion of frozen elements is too high.

 \begin{figure}[htb]
 \setbox41=\vbox{ \hsize=8cm
The formation of clusters also
leads to a more practical difficulty namely the fact that 
once ants are part of a cluster their spatial density
becomes so high that it is difficult to count them.
As they form several layers, counting becomes 
nearly impossible
even on high resolution pictures. Recently, we have tried
an alternative method which consists in weighing rather
than counting. This method is well suited 
for small beetles whose unit weight is of the order of
15mg (see Fig 3). 
It is more difficult for ants whose typical weight
(e.g. workers of ``Formica japonica'')
is about 3mg. It is altogether impossible for drosophila
whose typical weight is around 0.2mg.
}
\setbox42=\vbox{ \hsize=8cm
\vskip 1mm
\centerline{\psfig{width=4cm,figure=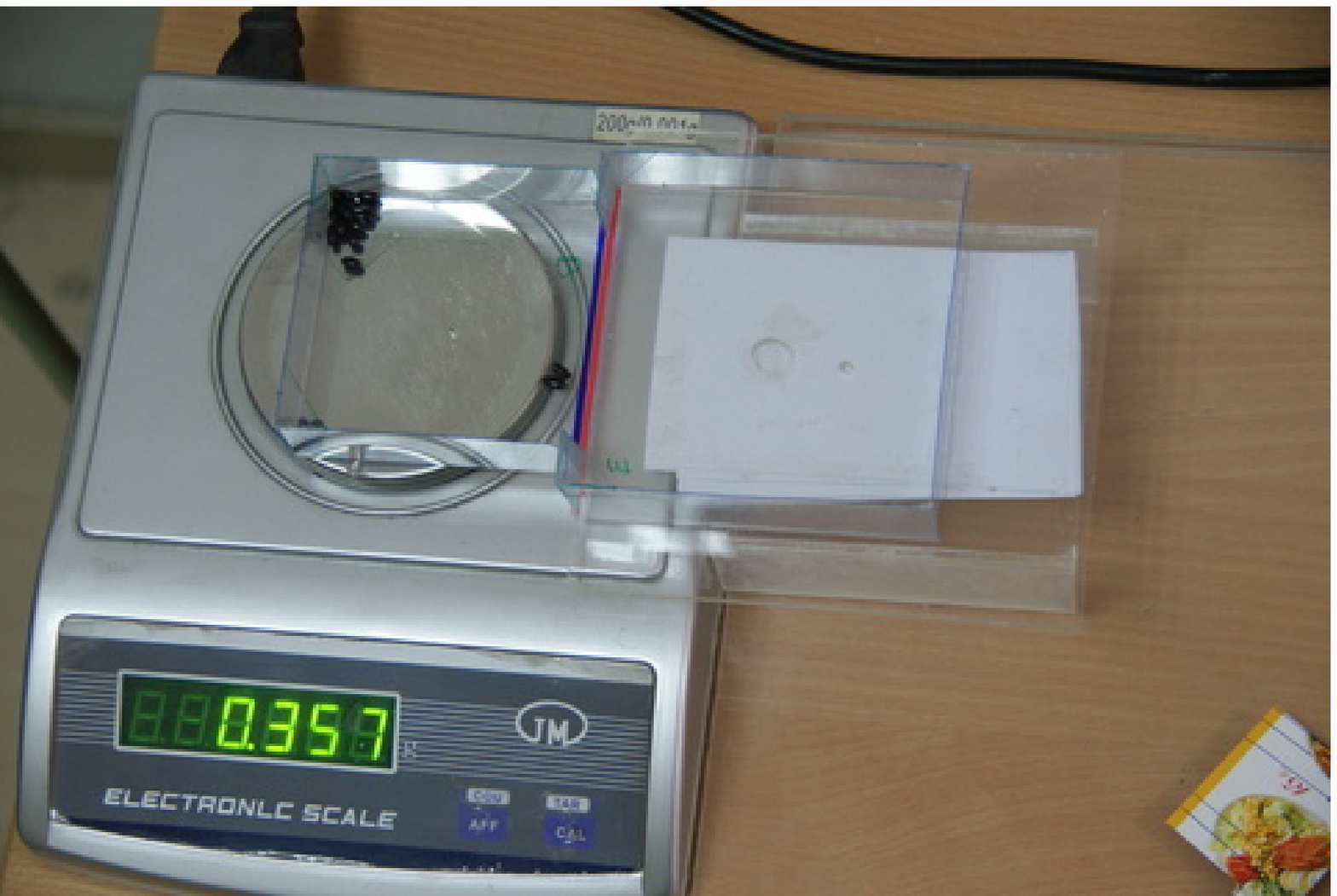}}
\qleg{Fig. 3: Container with weighing device on one side.}
{The compartments $ A $ and $ B $ 
are 
slightly (0.5mm) disjoined along the blue 
and red lines respectively
so that the weight measured by the scale
corresponds only to the beetles contained in the left-hand
side part but that the beetles can nevertheless cross 
from $ A $ to $ B $ and vice-versa.
Here most of the beetles have formed a cluster
in a corner. 
The weight is 357 mg which, when divided
by 15 mg, gives a total of 24 beetles.}
{}
}
\centerline{\hfill \hskip 10mm \box41  \quad \box42 \hfill}
%
\end{figure}

\qA{Consistency tests} 
For a liquid inter-molecular attraction can
be estimated through various means and variables: evaporation rate, 
equilibrium pressure of vapor, boiling temperature, heat of
vaporization. It is the fact that such estimates are 
(at least most often)
consistent with one another which gives us confidence in them.
One would like to do the same here. 
\qpar

A simple qualitative consistency test is provided by the
following ``evaporation'' experiment.
One takes a test tube containing some 50
drosophila and one makes them all move to the bottom of the tube
by hitting the tube on a table. Then, very quickly%
\qfoot{This movement must be fast because drosophila have a
natural tendency to go upward.}
one puts the tube
on the table in horizontal position. Let us assume that the bottom
of the tube is on the left.
After a few seconds, some 5 flies will have reached the
right-hand side, and may be 10 others will be in the middle of the tube.
If one waits 5mn, the flies will be distributed fairly uniformly
throughout the tube. 
\qpar

If one repeats the same experiment with ``Tenebrio molitor'' beetles
it will be seen that after 5mn almost all insects are still 
together on the left-hand side of the tube.
\qpar
This experiment can be repeated in a more precise way by
using the following procedure. 
\begin{figure}[htb]
\centerline{\psfig{width=10cm,figure=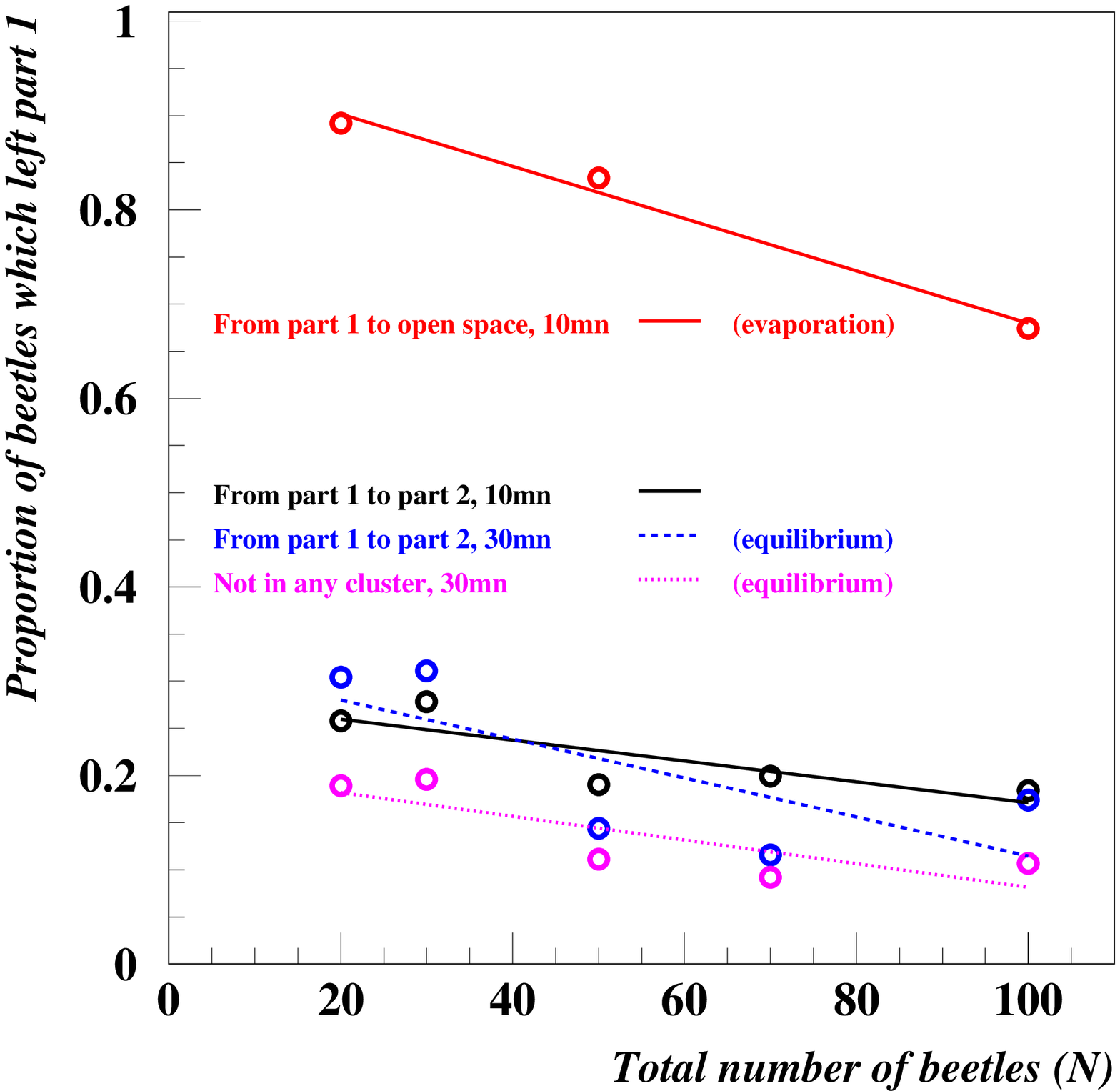}}
\qleg{Fig. 4a: ``Evaporation'' experiment with beetles.}
{In the evaporation version of the experiment (top) the beetles
move from the container into open space (i.e. the laboratory
table) whereas in the equilibrium version they move from
part 1 of the container into part 2 of same size. In the first
case almost no beetles come back into the container just like
the molecules in the evaporation of a liquid.
The graph shows that the dropout rate decreases when the 
size of the population increases pointing to greater
attraction power of larger groups. 
In physics similar effects can be observed. For instance
the vapor pressure around dropplets of
liquid decreases when the dropplets increase in size (Kelvin
equation) and the melting point of gold particles
increases with the diameter of the particles (Buffat and
Borel 1976, p. 2294).\qL
For each value of $ n $ the experiment was repeated 10 times,
which means that 80 experiments were performed altogether. 
For the 10 repetitions the coefficient of variation $ \sigma/m $
was around 50\%.
The slopes of the regression lines (with the numbers of beetles
expressed in thousands) 
are as follows (the error bars
correspond to a probability level of 0.90):\qL
evaporation: $ -2.8 \pm 0.5 $; 1 to 2, 10mn: $ -1.10\pm 0.7 $;
1 to 2, 30mn: $ -2.2\pm 1.8 $; not in cluster: $ -1.3\pm 0.8 $.
The average slope is $ a=-2.0 $.}
{The experiments were done in November 2012 by Ms. Mengying Feng and 
Shuying Lai from Beijing Normal University, Department of Systems
Science.}
\end{figure}
%
\begin{figure}[htb]
\centerline{\psfig{width=10cm,figure=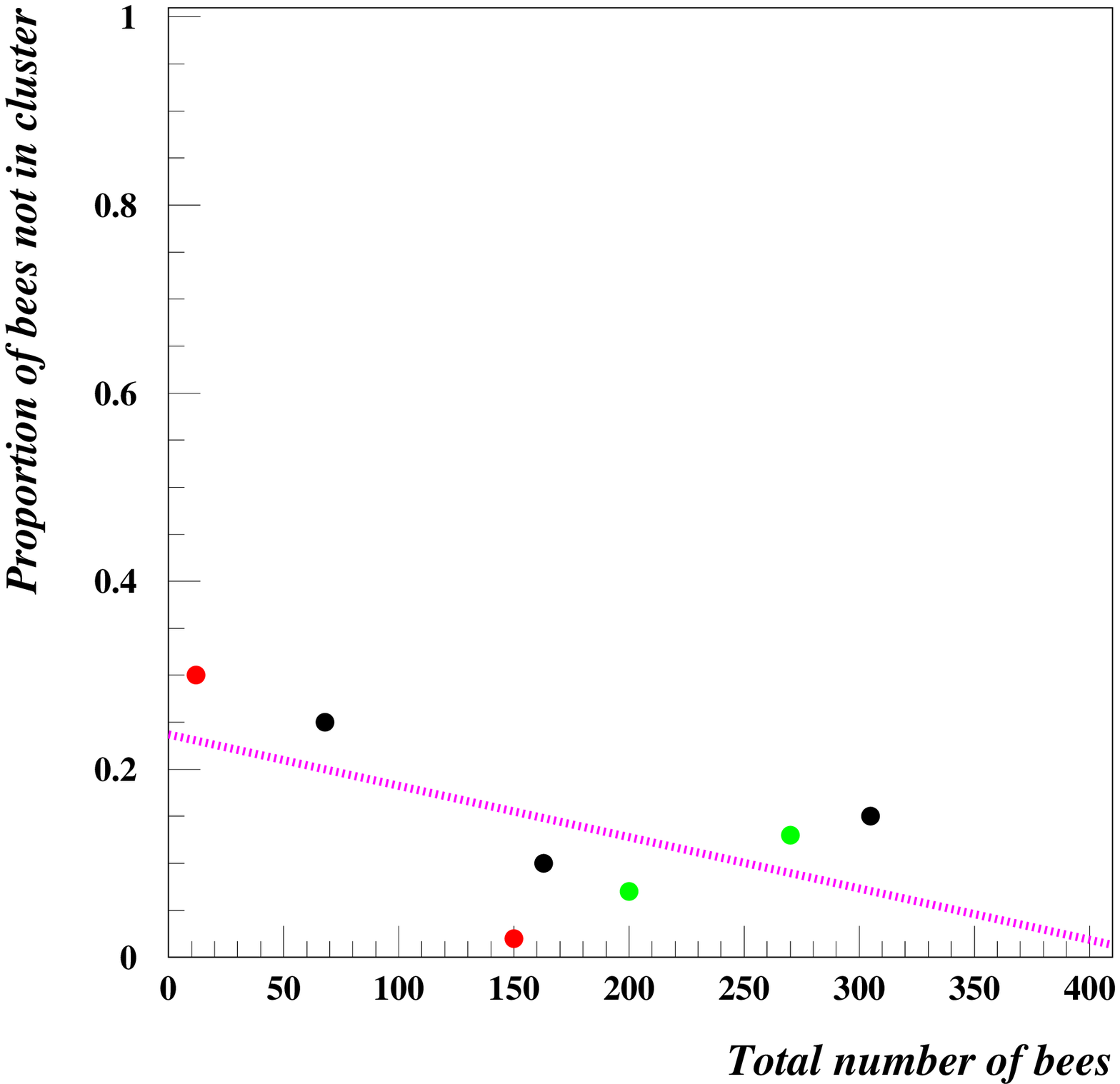}}
\qleg{Fig. 4b: ``Not in cluster'' experiment with bees.}
{After formation of a bee cluster, the number of those outside
of the cluster were counted. The duration of each experiment was 
comprised between one hour and one hour and a half. The three
different colors correspond to slightly different experimental
conditions. For instance, for the black data points there was
a single cluster whereas for
the red points two clusters formed. In the latter case
we divided all numbers by 2. 
The slope of the regression line (also expressed per 1,000 bees)
for the 7 experiments, 
namely $ a=-0.55\pm 0.68 $, is 2.3 times smaller than the 
``not in cluster'' slope in the beetles experiment.}
{The experiments were done in June and July 2012 by Mrs.
J. Darley and B. Roehner in Val Fleury (western suburb of Paris).
The bees were Appis Mellifera mellifera.}
\end{figure}

The experiment starts after a number $ n $ of beetles has been
introduced into a container that we will call part 1.
In the ``evaporation rate'' version
of the experiment, the beetles can just walk out into open space.
In the equilibrium version of the experiment the opening of part 1
leads to a container of same size. In this case, most often, the
beetles formed a cluster both in part 1 and in part 2. 
However not all the beetles were in the clusters.
This leads to the definition of
two different variables: $ n_2(t) $, the
number of beetles in part 2 at time $ t $, and $ n_3(t) $,
the number of beetles which are not in a cluster. It is this 
latter variable
which is the analog of the molecules in the vapor phase over a liquid.
The observations summarized in the figure show that whether in
the non-equilibrium case of evaporation or in the quasi-equilibrium
case, the escape rate decreases when the number of beetles increases.
A natural interpretation is that the combined attraction of
$ n $ beetles on one of them increases with $ n $%
\qfoot{More precisely, one can say that the experiment displays
two competing forces: (i) attraction and (ii) increased volatility.
The increased volatility likely comes with the beetles'
new environment. Indeed, when they are left alone for a long time
they cluster together instead of occupying the whole available area,
a typical liquid-like behavior.}
.

\qI{Conclusion}

In physics real progress occurs when there is a fruitful dialogue
between theory and observation. This is currently one of the problems
faced by string theory. There is a similar problem with computer
simulations of social phenomena because of the fact that they
rarely lead to testable predictions and when they do, most often,
the requested statistical data turn out to be unavailable. 
Thus, there is almost no dialogue between theory and observation.
This greatly hampers real progress. 
\qpar
For experiments on groups of insects the situation 
is completely different. Usually a model set up to account for
a given phenomenon also leads to predictions for other phenomena.
The nice feature is that most often the experiments required for
exploring those phenomena can be designed and implemented within
a few days. In terms of speed and convenience such experiments 
are very much alike computer simulations. In addition they
allow a real dialogue between theory and observation. 
\qpar

What kind of experiments should be tried by physicists?
Clearly, it would be useless to repeat the experiments already
done by entomologists. So far however, entomologists
have given only scant attention to the exploration of {\it group
effects}. The experiments which come closest to those that
we advocate are probably those conducted by the teams of
Jean-Louis Deneubourg (a former physicist)
in Brussels and Deborah Gordon at Stanford.
However, because they confined themselves to the study of ants both
Deneubourg and Gordon could not draw on the benefits and broader
perspective that might come from {\it comparative studies}%
\qfoot{A few centuries ago when physicists studied the phenomenon
of ``free'' fall they did not confine themselves to falling apples.
Indeed, comparative observation was the only way to demonstrate
that, at least in air, the
law is fairly independent of the shape and density of the falling
object. This was a milestone in the development of classical
mechanics.}%
.
\qpar

Appendix B gives some practical hints for performing experiments 
with living organisms. We hope that this information will
enable a number of other groups to carry out such experiments.
This is a field where there is much to explore. For instance,
some preliminary observations convinced us that 
the temperature%
\qfoot{In a general way observation shows that
the ``temperature'' of the living organisms
(in the sense of statistical physics) is determined by the 
temperature of the environment.}
plays a role in this kind of experiments which is 
fairly similar to what can be seen in
chemistry and statistical physics. However, this
must be confirmed and documented by a set of systematic experiments.

\vskip 4mm
{\bf Acknowledgments:} For their helpful comments we 
would like to thank the researchers who attended
a seminar given by one of us at the Department of Economics of
the University of Tokyo on 28 November 2012.\qL
One of the experiments reported above was done 
in Kunming at 
the Eastern Bee Institute of Yunnan Agricultural University, China;
we wish to express our appreciation to Prof. He and Tan
for their hospitality and to Dr. Chen and Wang for their help.\qL
We also want to express our gratitude to Ms. Mengying Feng and
Shuying Lai from the Department of Systems Science of Beijing Normal
University whose results were mentioned in our
discussion of the  ``evaporation'' rate of beetles.

\vfill\eject

\appendix

\qI{Appendix A: Variance of a sum of correlated variables}

\def\qcir#1{\hat #1}

{\normalsize

We proceed in several steps. 
\qpar
For the sake of simplicity we first
consider the average of a sum of three correlated
random variables $ X_1, X_2, X_3 $ of mean $ m $ 
 and identical standard deviation
$ \sigma $. Our objective is to compute the variance
of $ S_3=X_1+X_2 + X_3 $.
\qL
By definition of the variance 
$ \sigma^2(S_3) = E\left[ \left(S_3 -E(S_3)\right)^2 \right] $.
One knows that
the expectation of a sum of random variables is always equal to
the sum of the expectations, whether the variables are correlated
or not. Thus: $ E(S_3)=E(X_1)+E(X_2)+E(X_3) $.\qL
Consequently:
$$ \sigma^2(S_3) = E\left[ (\sum_{i=1}^3 \hat X_i)^2\right],\quad
\hbox{where: } \qcir{X_i}=X_i-E(X_i) $$

Thus,
$$ \sigma^2(S_3) =\sum_{i=1}^3 E({\qcir{X_i}}^2) + 
2\left[ E(\qcir{X_2}\qcir{X_3}) + 
E(\qcir{X_3}\qcir{X_1}) + E(\qcir{X_1}\qcir{X_2}) \right] $$

We express the expectations of the products by introducing the coefficient
of correlation of the $ X_i $:
$ r_{ij} = E(\qcir{X_i}\qcir{X_j})/\sigma^2  $.
Thus: $ \sigma^2(S_3)=3\sigma^2 + 2\sigma^2 (r_{23} + r_{31} + r_{12} ) $
\qpar

From that point on, we will consider two cases.

\qA{Uniform correlations}

Introducing the mean of the $ r_{ij} $, $ \overline r =
(r_{23}+r_{31}+ r_{12})/3 $, we obtain:
$$ \sigma^2(S_3)= 3\sigma^2[1+2\overline r] $$

This formula has an obvious generalization to an arbitrary 
number $ n $ of random variables:
$$ \sigma^2 (S_n) = n\sigma^2 g^2, \quad g^2=(n-1)\overline r +1    $$

where:
$$ \overline r = { 1 \over [n(n-1)/2] } \sum_{i\not = j}^n r_{ij} $$

\qA{Ising-like correlations}
For a one dimensional Ising spin system the correlation 
between spins $ i $ and $ j $ is: 
$ r_{ij}=\eta^{|i-j|} $ 
where $ \eta $ can be expressed (if one wish)
as a function of the parameters which
define the interaction between the spins (see Glauber 1963 p. 299,
formulas (56) and (57)).
\qpar

Introducing this expression of $ r_{ij} $ gives:
$ \sigma^2(S_3)=\sigma^2(3 + 2\eta + \eta^2) $

In extending this formula to any $ n $, one needs to
express the finite sum $ f(\eta)=\sum_{i=0}^{n-2}\eta^i $ 
(as well as its derivative $ f'(\eta) $).
Instead of using the exact expression
$ f(\eta)=(1-\eta^{n-1})/(1-\eta) $ we will consider that the term 
$ \eta^{n-1} $ is negligible with respect to 1, which means that
we approximate the finite sum by the corresponding infinite series.
This approximation is acceptable for our experiments because most
of the time $ n>20 $. Of course the approximation is no longer
valid when $ \eta \rightarrow 1 $ but $ \eta=1 $ is the case of
uniform correlation already considered above.
\qpar
Under this assumption one obtains finally:
$$ \sigma^2(S_n) =n\sigma^2\left[1+ { 2\eta\over 1-\eta }
\left(1-{ 1\over n(1-\eta) }\right)\right] $$ 
or:
 $$ g^2(n)= { \sigma^2(S_n)\over n\sigma^2 }=
{ 1+\eta\over 1-\eta } - { 2\eta\over n(1-\eta)^2 } $$

Due to the approximation made in the derivation,
this formula is not valid when $ n $ is 
close to 1. We have seen above that for $ n=2,3 $ one gets:
$$ g^2(2)=1+\eta,\quad g^2(3)=1+(4/3)\eta +(2/3)\eta^2 $$ 
which shows that the function
$ g^2(n) $
increases toward its asymptotic limit $ (1+\eta)/(1-\eta) $.
\qpar

{\bf Remark}\quad  
Can the Ising case be seen as a special instance
of the uniform case? Formally, it may seem so. However, the real
picture emerges when we consider large values of $ n $.
In the Ising case, due to the exponential decrease, all 
elements in the correlation matrix are almost equal
to zero except for a zone around the first diagonal
whose width depends only upon $ \eta $. Consequently, for 
such a matrix the 
average correlation goes to zero when $ n $ becomes 
larger. \qL
This observation shows three things. (i) It
would be irrelevant to treat the Ising case as a special
instance of the uniform case. (ii) The fact that in the Ising
case $ \overline r \simeq 0 $ helps to explain that
the ratio $ g^2(n) $ remains basically constant instead of
increasing. (iii) It explains why we used the
expression ``uniform
correlations'' to designate the first case. 
The correlations are uniform in the sense that when 
$ n\rightarrow \infty $ the number of elements of
the correlation matrix that are ``substantially''
different from zero must remain of the
same order of magnitude as $ n $. For a distance-dependent
correlation, this means that the decrease with distance
must be slow enough. 

\qA{Simulations}
So far we did not need to make the assumption that 
the $ X_i $ are
Bernoulli variables, that is to say variables
taking only the values $ 0 $ and $ 1 $
\qfoot{When $ P\{X=1\}=p $ such a variable will be noted 
as Ber($ p $).}%
. 
However, if one wishes
to carry out a simulation there is a convenient algorithm
which works only for Bernoulli variables (Lunn and Davies 1998).
The relevant formulas can be summarized as follows:
\qdec{ {\bf Simulation of
uniform correlations between $ n $ Bernoulli variables} \quad 
$ Z $ and $ Y_i $ are Ber($ p $) random variables while the
$ U_i $ are Ber($ \sqrt{r} $) random variables. 
Then, the variables $ X_i $ defined as:
$$ X_i=(1-U_i)Y_i+U_iZ,\quad  i=1,\ldots n $$
are correlated Bernoulli variables with the following properties:
 $$ E(X_i)=p,\quad E(X_i^2)=p,\quad \hbox{Cor}(X_i,X_j)=r,\quad  
i\not= j $$
}
\qpar
It can be noted that this algorithm works only for positive
correlations between the variables.
\qpar

\qdec{ {\bf Simulation of correlated Ising-like 
Bernoulli variables} \quad 
$ Y_i $ are Ber($ p $) random variables while the
$ U_i $ are Ber($ \eta $) random variables. 
Then, the variables $ X_i $ defined as:
$$ X_1=Y_1,\quad  X_i=(1-U_i)Y_i+U_iX_{i-1},\quad  2\le i\le n $$
are correlated Bernoulli variables with the following properties:
 $$ E(X_i)=p,\quad E(X_i^2)=p,\quad 
\hbox{Cor}(X_i,X_j)=\eta^{|j-i|},\quad  i\not= j $$
}
\qpar

Fig 1b presents a simulation of the trajectories of ants.
Such simulations are useful for testing the estimation procedure.
How were they done? 
\qbu First of all, in order to introduce
a time-continuity which obviously exists in real experiments
we generated the $ n $ random variables $ Z_{i,t} $
through $ n $ independent first-order auto-regressive
processes. 
\qbu Then, the correlations between the variables were
introduced following the so-called Cholesky procedure
by defining the $ X_{i,t} $ as appropriate linear combinations of the
$ Z_{j,t}, j=1,\ldots n $.

\qI{Appendix B: Experimental ``toolkit''}

Just in order to convince the reader that experiments with 
insects can be done fairly easily we give some practical hints.
It is indeed possible to do
this kind of experiments with fairly little sophisticated
equipment. \qL
Basically, the needs can be summarized as follows:
First one needs to get the {\it living organisms}.
\qbu Ants can be easily collected (at least in spring and summer)
by putting appropriate food as a bait on a Bristol board just 
a few centimeters away from the entrance of a colony.
Within one hour and depending on the species a few hundred
ants may gather on the Bristol board.
\qbu Drosophila can be obtained from biology laboratories.
\qbu Flies and beetles 
can be bought in the form of larvae (worms)
destined to fishermen or for feeding big aquarium-fishes. 
The waiting time between the larvae stage and the emergence of the
adults ranges from less than one week to a few months
depending on species, temperature and time of year.
\qpar

Secondly, in many cases, one needs a small bottle of
carbon dioxide to make the insect sleep in order to be able to handle
them easily. Carbon dioxide has an almost instantaneous
anesthetic effect on all these insects. According to a paper
published in the Journal of Experimental
biology (Ribbands 1950)
anesthesia through carbon dioxide does not
infer a memory loss and changes only slightly the behavior
of bees. It is probably safe to assume that the
effect on the other insects mentioned above is similar.
\qpar

Next one needs an appropriate container. A simple solution
is to cut it into a piece of flexible plastic (such as PVC)
of adequate thickness (3mm to 5mm is usually enough).
This is illustrated in Fig. 1a.
\qpar

Finally, one needs a counting device. Taking pictures
and counting by hand is a simple solution but not always
satisfactory especially for counting the elements in a
cluster. For this reason we have developed a weighing method
(illustrated in Fig. 3).
\qpar

Clustering phenomena also occur among bacteria and micro-organisms
that are present in so-called biofilms which 
form at the surface of liquids. 
Because of the small size and high
numbers of such elements one is in a situation
fairly similar to physical systems. For instance, it can
be mentioned that
inter-molecular forces such as
van der Waals forces play a significant role in the movements of
such micro-organisms. \qL
Studying the {\it collective behavior} of such  populations from
the perspective of physics seems a promising field. 
However, in contrast to the study of insects, it requires special
laboratory devices and equipment.
 
}   

\vskip 5mm
{\bf References} 
\qpar

\qparr
Aoki (M.), Yoshikawa (H.) 2007: Reconstructing macroeconomics:
a perspective from statistical physics and combinatorial
stochastic processes. Cambridge University Press, New York.

\qparr
Buffat (P.), Borel (J.-P.) 1976: Size effect on the
melting temperature of gold particles. 
Physical Review A, 13,2287-2298.

\qparr
Glauber (R.J.) 1963: Time-dependent statistics of the
Ising model. Journal of Mathematical Physics 4,294-307.

\qparr
Iyetomi (H.) 2012: Labor productivity distribution
with negative temperature. 
Progress of Theoretical Physics Supplement 194, 135-143.

\qparr
Iyetomi (H.), Nakayama (Y.), Aoyama (H.), Fujiwara (Y.), Ikeda (Y.),
Souma (W.) 2011: 
Fluctuation-dissipation theory of input-output interindustrial
relations.\qL
Physical Review E 83, 016103-1 to 016103-12.

\qparr
Lunn (A.D.), Davies (S.J.) 1998: A note on generating correlated
binary variables. Biometrika 85,2,487-490.

\qparr
Papoulis (A.) 1965: Probability, random variables and
stochastic processes.
McGraw-Hill Kogakusha, Tokyo.

\qparr
Ribbands (C.R.) 1950: Changes in the behavior of honeybees
following their recovery from anaesthesia.
Journal of Experimental Biology 27,3,302-309.

\qparr
Roehner (B.M.) 2004: A bridge between liquids and socio-economic
systems: the key-role of interaction strengths. 
Physica A 348, 659-682.

\qparr
Roehner (B.M.) 2012: Estimating interaction strength 
in populations of living organisms.
Working report, LPTHE, University Pierre and Marie Curie, Paris.
Available on the following website:\qL
http://www.lpthe.jussieu.fr/~roehner/effusion.pdf

\end{document}